\documentclass{article}
\usepackage{multirow}%
\usepackage{amsmath,amssymb,amsfonts}%
\usepackage{amsthm}%
\usepackage{mathrsfs}%
\usepackage[title]{appendix}%
\usepackage{xcolor}%
\usepackage{textcomp}%
\usepackage{manyfoot}%
\usepackage{booktabs}%
\usepackage{algorithm}%
\usepackage{algorithmicx}%
\usepackage{algpseudocode}%
\usepackage{listings}%

\newtheorem{lemma}{Lemma}

\newtheorem{remark}{Remark}%

\newcommand{\B}{\mathcal{B}}
\newcommand{\card}{\operatorname{card}}

\title
{
Maximum Likelihood Estimators of Quantum Probabilities
}

\author{Mirko Navara, Jan \v Sevic\\
Department of Cybernetics\\, Faculty of Electrical Engineering,\\ {Czech Technical University in Prague},\\
{Czech Republic},\\
\texttt{\{navara,sevicjan\}@fel.cvut.cz}}

\date{}

\begin{document}

\begin{abstract}
Classical probability theory is based on assumptions which are often violated in practice. Therefore quantum probability is a proposed alternative not only in quantum physics, but also in other sciences. However, so far it mostly criticizes the classical approach, but does not suggest a working alternative. Maximum likelihood estimators were given very low attention in this context. We show that they can be correctly defined and their computation in closed form is feasible at least in some cases.
\end{abstract}



\maketitle


\section{Motivation of quantum probability}

Classical probability is based on the following principles:
\begin{enumerate}
 \item Results of experiments are evaluated in yes-no terms.
 \item Experiments can be repeated arbitrarily many times.
 \item The initial state of a random experiment can be reconstructed. 
\end{enumerate}
When the first principle is violated, we need probability of many-valued statements; this is not the direction studied in this paper.
The second principle is theoretical; we apply the theory also to experiments whose repeatability is limited by some finite number, 
e.g., the number of people in the world.
Here we concentrate on the case when the third principle does not hold.
Quantum measurements cause irreversible changes of the system and the initial state can never be reconstructed.
This phenomenon is typical in quantum physics, therefore we speak of quantum probability, 
but it occurs in many other situations. E.g., elections cannot be repeated under the same conditions,
so we have no tool for a direct comparison of two strategies---only one of them can be tested.
Such situations are frequent in sociology, psychology, and recently even in machine learning;
we work with such big databases that we cannot keep them all in the memory and each experiment will be performed in different circumstances (e.g., finding different random data on the internet).
For simplicity, we speak here of quantum events.

When we apply the classical Kolmogorovian probability to such systems, we obviously make an error.
It is not the statistical error, well described by the theory, but an unpredictable error due to the use of assumptions which are not satisfied. The only correct way is to use more general theory fitting to the real situations, and this is quantum probability.
It shows why the classical results may be erroneous.
However, it is easy to criticize, but difficult to develop a working alternative.
In particular, it is hard to introduce conditional probability of fuzzy events; 
performing one measurement, the Heisenberg's uncertainty principle makes some other measurements impossible,
although they were allowed at the beginning and therefore must be included in the probabilistic model.
Many statistical tools are based on the laws of large numbers and the existence of infinite sequences of independent random experiments,
and we do not see a way how to generalize them to quantum probability.

However, the second principle seems to be valid in quantum probability at least to the extent that a large repetition of experiments 
can give us useful data. Although the initial conditions are not guaranteed, their partial knowledge can be described by probabilities.
The possibility of acquiring large data enables to use them for estimation of parameters of the model.
Here we show that unknown probabilities can be estimated by a maximum likelihood estimator (MLE) generalized to quantum probabilities.
As far as we know, the only such attempt was made in
\cite{KlayFoulis}.
It solved several particular cases and demonstrated that the results obtained by quantum probability substantially differ from those offered by the classical theory.
We continue this research in a~similar direction.

\section{Motivating example}\label{s:motivating}

Let us demonstrate the situation on a simple example.
A~coach of a tennis team decides which of two players, A or B, will play the first match.
(Many random inputs influence the result, the choice of the oponent, health conditions, etc.
We hope that they can be described by the probabilities of outcomes and will still allow to compare the players on the base of long-term data.)
There are five possible outcomes:
\begin{enumerate}
\item $a$ ... A played and won,
\item $b$ ... B played and won,
\item $c$ ... A played and lost,
\item $d$ ... B played and lost,
\item $e$ ... the match was cancelled.
\end{enumerate}
When A is nominated, the possible outcomes are $a$, $c$, and~$e$,
with the respective probabilities $p(a)$, $p(c)$, and~$p(e)$, summing up to~$1$.
When B is nominated, the possible outcomes are $b$, $d$, and~$e$,
with the respective probabilities $p(b)$, $p(d)$, and~$p(e)$, summing up to~$1$.
The coach can decide between two random experiments, each described by a classical probability.
However, only one of the two random experiments can be performed, disabling the other.

The two possible experiments share one common result,~$e$.
While the coach knows which player he nominated,
an independent observer knows only which of the five outcomes occurred.
In case~$e$, when the match was cancelled, it is not known which of the two experiments was performed.
The observer knows only a long series of outcomes of preceding matches.
Based on it, he wants to estimate non-negative probabilities 
$p(a)$, $p(b)$, $p(c)$, $p(d)$, and~$p(e)$, subject to restrictions
\begin{equation}\label{eq.motivating}
p(a)+p(c)+p(e)=p(b)+p(d)+p(e)=1\,.
\end{equation}
We shall show that this may be solved by an easy modification of a classical maximum likelihood estimator,
which has also an intuitive interpretation.
However, other models lead to more complicated tasks.
We solve them at least for some classes of quantum probability models.

\section{Basic notions}

We shall deal only with finite models.
That is reasonable because the estimator uses finitely many results and cannot say anything about those which did not occur in the sample.

As event structures, generalizing classical Boolean algebras, we use orthomodular lattices, or even orthomodular posets
\cite{Beran, Kalmbach}.
To make the text understandable also to readers not well acquainted with these algebraic structures, 
we start from their hypergraph representation in simple cases and leave more general ones to the end of the paper.

A \emph{hypergraph} 
is a couple $(A,\B)$, where $A$ is a non-empty set and $\B$ is its covering by non-empty subsets.
The elements of $A$ are \emph{vertices} and represent all possible 
\emph{outcomes, results, atoms}, or
\emph{answers} of the system.
The elements of $\B$ (some subsets of~$A$) are \emph{edges} and represent all possible 
\emph{operations, experiments, blocks}, or
\emph{questions}.
Here we follow the terminology of 
\cite{KlayFoulis}
and call the elements of $A$
\emph{outcomes}
and the elements of $\B$
\emph{operations}.%
\footnote{Only if there is a risk of confusion, we speak of edges rather than operations.}
An operation $B\in\B$ is the set of all its possible outcomes.

\begin{figure}[ht]
\begin{center}
\unitlength=1mm
\special{em:linewidth 0.4pt}
\begin{picture}(60,50)(0,5)
\put(10  ,10  ){\circle*{2}}
\put(50  ,10  ){\circle*{2}}
\put(20.5,27.5){\circle*{2}}
\put(39.5,27.5){\circle*{2}}
\put(30  ,43.3){\circle*{2}}
%
\put(10.5,10.8){\line(3,5){19  }}
\put(49.5,10.8){\line(-3,5){19  }}
\put(29,45){$e$}
\put(7.5,7){$a$}
\put(41,28){$d$}
\put(51,7){$b$}
\put(17,28){$c$}
\put(47,18){$B$}
\put(11,18){$A$}
\end{picture}
\caption{Greechie diagram of the motivating example}
\label{f:MO2x2}
\end{center}
\end{figure}
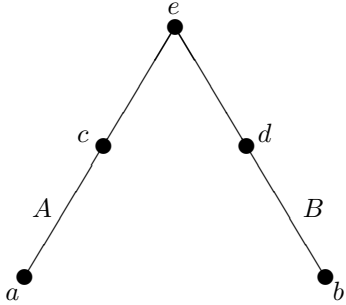

In the example from Section~\ref{s:motivating},
$A=\{a,b,c,d,e\}$, 
$\B=\{\{a,c,e\}, \{b,d,e\}\}$.
The respective hypergraph is drawn in Fig.~\ref{f:MO2x2}.

We see that a probability is uniquely described by a non-negative evaluation of the outcomes
which sums up to~$1$ for each operation.
Formally:
$p\colon A\to[0,1]$ such that
\begin{equation}\label{sum}
\forall B\in\B: \sum_{x\in B} p(x) = 1\,.
\end{equation}
Such a probability assignment can be uniquely extended to other events which combine the (elementary) outcomes by logical operations,
e.g., \begin{equation}
p(a\lor c)=p(a)+p(c)=1-p(e)=p(e')\,,
\end{equation}
where $e'$ denotes the negation of~$e$.
Notice that also
\begin{equation}
p(b\lor d)=p(b)+p(d)=1-p(e)=p(e')\,,
\end{equation}
thus
\begin{equation}
p(a\lor c)=p(b\lor d)\,.
\end{equation}
This is not only an equality of probabilities; $a\lor c$ and $b\lor d$ represent the same event,
``the match was not cancelled''.
It has a non-unique representation as a disjunction of (elementary) outcomes.
This is what motivates the notion of orthomodular lattices for description of \emph{all} events of the system.
However, it is sufficient to determine the probability of the elementary events, called outcomes here.

Not all finite hypergraphs represent quantum operations and outcomes.
For that, necessary and sufficient conditions were proved in
\cite{Dichtl, Kalmbach}.
We do not quote these complicated results here because we will manage with some simple situations.
E.g., if every couple of operations shares at most one common outcome, then the necessary and sufficient conditions for a hypergraph to represent an orthomodular poset (resp.\ an orthomodular lattice) reduce to the conjunction of the following:
\begin{enumerate}
\item[(G1)] $\forall B_1,B_2\in\B, B_1\ne B_2: \card(B_1\setminus B_2)\ge2\,,$
\item[(G2)] the length of every cycle in $(A,\B)$ is at least~$4$ (resp.~$5$).%
\footnote{The notion of a cycle in a hypergraph is quite natural, though mathematically complicated. We refer to
\cite{Dichtl, Kalmbach}
for details.}
\end{enumerate}
The finite hypergraphs which represent orthomodular posets are called \emph{Greechie diagrams}.
They were introduced in
\cite{Greechie}.

\begin{remark}\label{rem:single}
 The two-element Boolean algebra corresponds to the Greechie diagram with a single vertex and a single edge, $(\{e\}, \{\{e\}\})$.
\end{remark}

In
\cite{KlayFoulis},
a more general approach was studied, based on so-called 
\emph{test spaces}, introduced by Foulis and Randall.
They admit also strict inclusion between Boolean algebras of events generated by outcomes of two different operations.
This allows to represent partial answers to questions, like \textit{``I do not tell you whom I vote for, but it definitely will not be X.Y.%
''}.
Here we restrict attention to the case when all events form an orthomodular poset and questions are maximal sets of mutually exclusive outcomes.

\section{Notation and formulation of the task}

%
%

We assume that we know the model of events described by a Greechie diagram $(A,\B)$.
Further, we have a sequence of realizations of outcomes.
Their order is unimportant, the only useful inputs are absolute frequencies $n(x)\in\mathbb N$ for all $x\in A$.
From these, we want to estimate the probabilities $p(x)\in[0,1]$, $x\in A$.
They have to satisfy the restrictions~(\ref{sum}).

We want to choose such $p\colon A\to[0,1]$ which maximizes the likelihood,
\begin{equation}\label{eq.likelihood}
L(p)=\prod_{x\in A} p(x)^{n(x)}\,.
\end{equation}
The maximum likelihood is positive and then
we may equivalently maximize the log-likelihood
\begin{equation*}
\ell(p)=\ln L(p)=\sum_{x\in A^+} n(x)\,\ln p(x)\,,
\end{equation*}
where the sum is taken over the set of all outcomes $x$ with non-zero probability, $A^+=\{x\in A \mid p(x)\ne0\}$.
If the absolute frequency $n(x)=0$ for some outcome~$x$, its probability $p(x)$ will not occur in the formula for likelihood.
Maximal likelihood is achieved for $p(x)=0$ (leaving maximum freedom for the probabilities of other outcomes) and we may omit such outcomes in our analysis.%
\footnote{We really achieve \emph{maximum} likelihood for $p(x)=0$. This does not appear in (\ref{eq.likelihood}); more exactly, it gives rise to a factor $p(x)^{n(x)}=0^0=1$ which does not affect the product. The sum of other probabilities is upper bounded by $1-p(x)$, which becomes the maximal bound,~$1$, and thus allows their maximal values.}
From now on, we shall assume that all outcomes have non-zero frequencies.%
\footnote{There is one significant difference: without such outcomes, condition (G1) may be violated.
However, this is only a technical assumption needed for the Greechie diagram to represent an orthomodular poset.
In order to have a hypergraph which properly represents the probabilities, this is not important and our computations work without this condition, too.}

In a classical system, $\B$ contains only one operation, which is~$A$, the Greechie diagram is $(A,\B)=(A,\{A\})$.
Then 
(\ref{sum}) attains the form $\sum_{x\in A} p(x)=1$
and MLE has a well-known solution, which is the empirical distribution with
\begin{equation*}
p(x)=\frac{n(x)} {\sum\limits_{y\in A} n(y)} \quad\text{ for all }\quad x\in A\,.
\end{equation*}

We shall use this principle several times, thus we formulate it as a simple lemma:

\begin{lemma}\label{lemma}
Given numbers $n_1,\ldots,n_k\in[0,\infty]$, we look for probabilities $p_1,\ldots,p_k\in[0,1]$ such that $\sum_i p_i=1$ and
\begin{equation*}
\sum_i n_i\,\ln p_i
\end{equation*}
is maximal.
This task has a unique solution, which satisfies
\begin{equation*}
p_i=\frac{n_i} {\sum\limits_j n_j}
\end{equation*}
for all~$i$;
equivalently, the ratio
\begin{equation*}
\frac{n_i} {p_i}=\sum_j n_j
\end{equation*}
is the same for all~$i$.
\end{lemma}

\section{Horizontal sums of classical models}\label{s:horizontal_BAs}

Assume now that the event system is described by a Greechie diagram $(A,\B)$ composed of $k$ \emph{mutually disjoint} operations $A_1,\ldots,A_k$, i.e.,
$A=\bigcup_i A_i$ and $\B=\{A_1,\ldots,A_k\}$.
Then $(A,\B)$ is called a~\emph{horizontal sum} of the classical systems described by the Greechie diagrams 
$(A_1,\{A_1\}),\ldots,(A_k,\{A_k\})$.
In this case, each outcome $x\in A$ is contained in a \emph{unique} operation $A_i$, $i\in\{1,\ldots,k\}$.

From the absolute frequencies 
$n\colon A\to\mathbb N$, we know that operation $A_i$ produced
$n(A_i):=\sum_{x\in A_i} n(x)$ outcomes.
Their probabilities can be estimated like probabilities conditioned by the fact that $A_i$ was the chosen operation.
The log-likelihood becomes
\begin{align}
\ell(p)&=\sum_{x\in A
} n(x)\,\ln p(x)
=\sum_i \sum_{x\in A_i
} n(x)\,\ln p(x) \nonumber\\
&=\sum_i n(A_i)\, \underbrace{\sum_{x\in A_i
} \frac{n(x)} {n(A_i)}\,\ln p(x)}_{\ell(p|A_i)}\,. \label{log-likelihood_horizontal}
\end{align}
The last sum can be understood as the log-likelihood of a conditional probability $p|A_i\colon A_i\to[0,1]$,
whose values on $A_i$ sum up to~$1$.
This log-likelihood is based on the subsample of outcomes belonging to~$A_i$.
Lemma~\ref{lemma} is applicable and the $i$th summand is maximized if the restriction $p|A_i$
\footnote{Here $p|A_i$ denotes the restriction of $p$ to~$A_i$ like a conditional probability; it may be understood this way, too.}
coincides with the coefficients 
$\frac{n(x)} {n(A_i)}$.
This maximization is independent of all other summands of the outer sum, thus giving a global maximum by
\begin{equation}\label{s:horizontal_BAs_max}
p(x)=\frac{n(x)} {n(A_i)}=(p|A_i)(x)
\end{equation}
for all $x\in A_i$ and all $i\in\{1,\ldots,k\}$.
Notice that the ratio
\begin{equation}\label{s:horizontal_BAs_ratio}
\frac{n(x)} {p(x)}=n(A_i)
\end{equation}
depends only on~$i$, not on the choice of $x\in A_i$.

This result is not surprising and is supported by intuition.
We shall generalize it in Section~\ref{s:constructible}.

\section{Motivating example continued}\label{s:motivating_cont}

The example from Section~\ref{s:motivating} is not a horizontal sum, outcome $e$ belongs to two operations
and has just one probability, summing the cases when it originates from the first or the second operation.
We shall demonstrate the MLE for this structure and generalize the procedure in the next section.

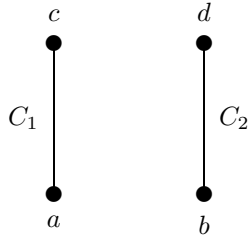
\begin{figure}[ht]
\begin{center}
\unitlength=1mm
\special{em:linewidth 0.4pt}
\linethickness{0.4pt}
\begin{picture}(40,40)
\put(10,10){\circle*{2}}
\put(10,30){\circle*{2}}
\put(30,10){\circle*{2}}
\put(30,30){\circle*{2}}
\put(10,10){\line(0,1){20}}
\put(30,10){\line(0,1){20}}
\put(10,7){\makebox(0,0)[ct]{$a$}}
\put(30,7){\makebox(0,0)[ct]{$b$}}
\put(10,33){\makebox(0,0)[cb]{$c$}}
\put(30,33){\makebox(0,0)[cb]{$d$}}
\put(8,20){\makebox(0,0)[rc]{$C_1$}}
\put(32,20){\makebox(0,0)[lc]{$C_2$}}
\end{picture}
\caption{Greechie diagram of a part of the motivating example}
\label{f:MO2}
\end{center}
\end{figure}

MLE requires to maximize the log-likelihood
\begin{equation*}
 \sum_{x\in\{a,b,c,d,e\}
} n(x)\,\ln p(x)
\end{equation*}
under restrictions~(\ref{eq.motivating}), which can be solved by standard techniques.
Another idea is to estimate first
\begin{equation}\label{eq.p(e)}
p(e)=\frac{n(e)} {\sum_{x\in\{a,b,c,d,e\}} n(x)}\,.
\end{equation}
The remaining outputs form the set $C=\{a,b,c,d\}$ and have a total probability $p(C):=1-p(e)$.
Probabilities of outputs from $C$ are restricted by conditions
\begin{equation*}
p(a)+p(c)=p(b)+p(d)=1-p(e)=p(C)\,.
\end{equation*}
If we define $
p(.|C)\colon C\to[0,1]$ by%
\footnote{Here $
p(.|C)$ is really an ordinary conditional probability, not a~restriction as in Section~\ref{s:horizontal_BAs}.}
$
p(x|C)=\frac{p(x)} {p(C)}$,%
\begin{equation}\label{eq.motivating3}
p(a|C)+p(c|C)=p(b|C)+p(d|C)=1\,.
\end{equation}
Maximization of the expression
\begin{equation}\label{eq.motivatingB}
\sum_{x\in C
} n(x)\,\ln p(x)
\end{equation}
under conditions (\ref{eq.motivating3})
is the same as MLE for the system described by the Greechie diagram $(C, \{C_1,C_2\})$, 
where $C_1=\{a,c\}$, $C_2=\{b,d\}$, 
see Fig.~\ref{f:MO2}.
This is the horizontal sum of two classical systems corresponding to operations $C_1,C_2$.
The MLE estimator of probability $
p(.|C)$ on $(C, \{C_1,C_2\})$ is based on the subsample from~$C$, excluding~$e$, which is of sample size $n(C):=\sum_{x\in C} n(x)$.
Similarly, we define $n(C_1):=\sum_{x\in C_1} n(x) = n(a)+n(c)$ and $n(C_2):=\sum_{x\in C_2} n(x) = n(b)+n(d)$.
According to Section~\ref{s:horizontal_BAs}, the maximum of (\ref{eq.motivatingB}) is achieved for
\begin{align*}
p(a|C)&= \frac{n(a)} {n(C_1)}=\frac{n(a)} {n(a)+n(c)}\,, \\
p(c|C)&= \frac{n(c)} {n(C_1)}=\frac{n(c)} {n(a)+n(c)}\,, \\
p(b|C)&= \frac{n(b)} {n(C_2)}=\frac{n(b)} {n(b)+n(d)}\,, \\
p(d|C)&= \frac{n(d)} {n(C_2)}=\frac{n(d)} {n(b)+n(d)}\,.
\end{align*}
This optimum is independent of the estimated value of $p(e)$.
A routine verification confirms that the choice (\ref{eq.p(e)}) is really optimal.
The remaining estimated probabilities are
\begin{align*}
p(a)&=p(C)\,p(a|C)=(1-p(e))\,\frac{n(a)} {n(a)+n(c)}\,, \\
p(c)&=p(C)\,p(c|C)= (1-p(e))\,\frac{n(c)} {n(a)+n(c)}\,, \\
p(b)&=p(C)\,p(b|C)= (1-p(e))\,\frac{n(b)} {n(b)+n(d)}\,, \\
p(d)&=p(C)\,p(d|C)= (1-p(e))\,\frac{n(d)} {n(b)+n(d)}\,.
\end{align*}
Notice that
\begin{align*}
\frac{n(a)} {p(a)}&=
\frac{n(c)} {p(c)}\,, &
\frac{n(b)} {p(b)}&=
\frac{n(d)} {p(d)}\,.
\end{align*}

We shall generalize this procedure in the next section.

\section{Products}\label{s:products}

The general construction of products (sometimes called direct sums) of algebraic structures is applicable to orthomodular posets and lattices. When expressed in terms of their Greechie diagrams, its description is cumbersome, but the understanding of probabilities on a product will be easy. It acts as follows:

Let $(B_1,\B_1),\ldots,(B_k,\B_k)$ be Greechie diagrams with $B_1,\ldots,B_k$ mutually disjoint.
Their \emph{product} is the Greechie diagram $(B,\B)$, where
\begin{align*}
 B&:=\bigcup_j B_j\,,\\
 \B&:=\biggl\{ \bigcup_j C_j \mid \forall j\in\{1,\ldots,k\}: C_j\in\B_j \biggr\}\,.
\end{align*}
Each choice $(C_1,\ldots,C_k)\in \B_1\times\ldots\times\B_k$ corresponds to one edge (operation) of~$\B$, 
restricting the probability $p$ on $\B$ by the equation
\begin{equation*}
\sum_{x\in\bigcup_j C_j} p(x)
=\sum_j \sum_{x\in C_j} p(x) =1\,.
\end{equation*}
If choices \mbox{$(C_1,C_2,\ldots,C_k), (C_1^*,C_2,\ldots,C_k)\in\B_1\times\ldots\times\B_k$} differ only in the first item, we see that
\begin{align}
 \sum_{x\in C_1} p(x) + \sum_{j=2}^k \sum_{x\in C_j} p(x) &=1
=\sum_{x\in C_1^*} p(x) + \sum_{j=2}^k \sum_{x\in C_j} p(x)\,,\nonumber\\
\sum_{x\in C_1} p(x)
&=\sum_{x\in C_1^*} p(x) \,. \label{eq:C_1}
\end{align}
Thus the value
\begin{equation*}
p(B_1):=\sum_{x\in C_1} p(x)\,,
\end{equation*}
is independent of the choice of $C_1\in\B_1$ and $p(B_1)$ is correctly defined.
Similarly, for any index $i\in\{1,\ldots,k\}$ (replacing index~$1$ in (\ref{eq:C_1})), we may correctly define the value
\begin{equation*}
p(B_i):=\sum_{x\in C_i} p(x)\,,
\end{equation*}
where $C_i\in \B_i$ can be chosen arbitrarily.
It is the probability that the experiment gives an output from~$B_i$.%
\footnote{Notice that we cannot sum over all $B_i$ if it has more edges (operations).}
We also define conditional probability $p(.|B_i)\colon B_i\to[0,1]$,
\begin{equation*}
p(x|B_i):=\frac{p(x)} {p(B_i)}
\end{equation*}
for all $x\in B_i$.

The log-likelihood can be expressed as
\begin{align*}
 \ell(p)
 &=\sum_{x\in B
} n(x)\,\ln p(x)
 =\sum_i \sum_{x\in B_i
} n(x)\,\ln p(x) \\
 &=\sum_i \sum_{x\in B_i
} n(x)\,\ln p(x|B_i) + \sum_i \sum_{x\in B_i
} n(x)\,\ln p(B_i) \,.
\end{align*}
The latter sum can be simplified to
\begin{equation*}
  \sum_i \sum_{x\in B_i
} n(x)\,\ln p(B_i) 
=\sum_i n(B_i)\,\ln p(B_i) \,,
\end{equation*}
where $n(B_i)
:=\sum_{x\in B_i
} n(x)=\sum_{x\in B_i} n(x)$ is the number of samples belonging to~$B_i$.
Lemma~\ref{lemma} proves that this expression achieves its maximum for
\begin{equation}\label{eq.product}
p(B_i)=\frac{n(B_i)} {\sum_j n(B_j)}\,.
\end{equation}

It remains to maximize the sum
\begin{equation*}
\sum_i \sum_{x\in B_i
} n(x)\,\ln p(x|B_i) \,.
\end{equation*}
Parameters describing its maximum are independent of the unconditional probabilities $p(B_i)$
and also independent between different indices~$i$.
Thus the task of MLE can be decomposed to the following steps:
\begin{enumerate}
\item Compute $p(B_1),\ldots,p(B_k)$ from~(\ref{eq.product}).
\item For each $i$, find MLE of the conditional probability $p(x|B_i)$. Using $p(B_i)$ from Step~1, convert it to an unconditional probability $p(x)=p(B_i)\,p(x|B_i)$, \mbox{$x\in B_i$}.
\end{enumerate}
Thus the task of MLE is decomposed to simpler subtasks for each factor separately.

The Greechie diagram from Section~\ref{s:motivating} can be considered a product of one one-element edge (representing the two-element Boolean algebra, see Remark~\ref{rem:single}) and a horizontal sum with two two-element edges.
Its MLE described in  Section~\ref{s:motivating_cont} is a special case of the procedure described here.

\section{Horizontal sums of quantum models}\label{s:horizontal}

Let $(A_1,\B_1),\ldots,(A_k,\B_k)$ be Greechie diagrams with $A_1,\ldots,A_k$ mutually disjoint.
We define
$A=\bigcup_i A_i$ and $\B=\bigcup_i \B_i$.
Then $(A,\B)$ is a Greechie diagram of an event structure called a~\emph{horizontal sum} of $(A_1,\B_1),\ldots,(A_k,\B_k)$.
As in the special case studied in Section~\ref{s:horizontal_BAs}, each outcome $x\in A$ is contained in a \emph{unique} summand $A_i$, $i\in\{1,\ldots,k\}$.

A probability $p$ on $(A,\B)$ is uniquely determined by its restrictions $p|A_i$, $i\in\{1,\ldots,k\}$, and these can be arbitrary probabilities on the horizontal summands $(A_i,\B_i)$.

The situation is analogous to that of Section~\ref{s:horizontal_BAs}.
From the absolute frequencies 
$n\colon A\to\mathbb N$, we know that the summand $A_i$ produced
$n(A_i):=\sum_{x\in A_i} n(x)$ outcomes.
Their probabilities can be estimated like probabilities conditioned by the fact that $A_i$ was the respective operation.
The log-likelihood is again expressed by (\ref{log-likelihood_horizontal}) and achieves its maximum for (\ref{s:horizontal_BAs_max}).
Thus the task of MLE is decomposed to simpler subtasks for each horizontal summand separately.

\section{Constructible lattices}\label{s:constructible}

Janowitz
\cite{Janowitz}
introduced \emph{constructible lattices} as the class generated from Boolean algebras 
by the operations of a product and a horizontal sum (applied arbitrarily many times).
This class does not contain all orthomodular lattices, but it played a role in numerous papers on quantum structures.
The example from Section~\ref{s:motivating} is constructible.

MLE works for horizontal sums and also for products, as described in Sections \ref{s:products} and~\ref{s:horizontal}.
Thus we directly obtain MLE for all finite constructible lattices,
just by subsequent application of these techniques.

\section{Non-constructible lattices
}

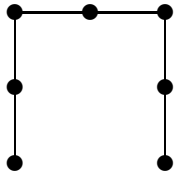
\begin{figure}[ht]
\begin{center}
\unitlength=1mm
\special{em:linewidth 0.4pt}
\linethickness{0.4pt}
\begin{picture}(40,32)(0,5)
\put(30,30){\circle*{2}}
\put(30,10){\circle*{2}}
\put(10,10){\circle*{2}}
\put(10,30){\circle*{2}}
\put(10,20){\circle*{2}}
\put(20,30){\circle*{2}}
\put(30,20){\circle*{2}}
\put(30,10){\line(0,1){20}}
\put(30,30){\line(-1,0){20}}
\put(10,30){\line(0,-1){20}}
\end{picture}
\caption{Greechie diagram of the smallest non-constructible orthomodular lattice}
\label{f:Pi}
\end{center}
\end{figure}

The simplest orthomodular lattice which is not constructible is described by its Greechie diagram in Fig.~\ref{f:Pi}.

\begin{figure}[ht]
\begin{center}
\unitlength=1mm
\special{em:linewidth 0.4pt}
\linethickness{0.4pt}
\begin{picture}(40,32)(0,5)
\put(30,30){\circle*{2}}
\put(30,10){\circle*{2}}
\put(10,10){\circle*{2}}
\put(10,30){\circle*{2}}
\put(10,21){\makebox(0,0)[cc]{$\vdots$}}
\put(20,30){\makebox(0,0)[cc]{$\cdots$}}
\put(30,21){\makebox(0,0)[cc]{$\vdots$}}
\put(30,10){\line(0,1){6}}
\put(30,30){\line(-1,0){6}}
\put(30,30){\line(0,-1){6}}
\put(10,30){\line(1,0){6}}
\put(10,30){\line(0,-1){6}}
\put(10,10){\line(0,1){6}}
\put(32,15){\makebox(0,0)[lc]{$A_3$}}
\put(8,15){\makebox(0,0)[rc]{$A_1$}}
\put(25,32){\makebox(0,0)[cb]{$A_2$}}
\put(32,32){\makebox(0,0)[lb]{$y_2$}}
\put(8,32){\makebox(0,0)[rb]{$y_1$}}
\end{picture}
\caption{Greechie diagram of a non-constructible orthomodular lattice for which MLE is feasible}
\label{f:Pi2}
\end{center}
\end{figure}
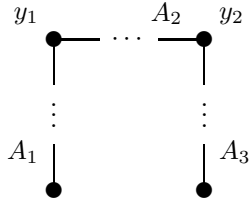

For it, the preceding ideas do not work and we do not see any intuitive solution.
We shall show that the MLE is still feasible for it.
More generally, we present it for
a Greechie diagram with edges (operations) $A_1$, $A_2$, $A_3$ satisfying
$A_1\cap A_2=\{y_1\}$,
$A_2\cap A_3=\{y_2\}$,
$A_1\cap A_3=\varnothing$.
They form the same shape, but may have more than three outcomes of each operation, see Fig.~\ref{f:Pi2}.

Like Sections \ref{s:products} and~\ref{s:horizontal}, the solution can be done in two steps.

First, we define:
\begin{align*}
B_1 &= A_1 \setminus A_2= A_1 \setminus \{y_1\} \, , \\
B_2 &= A_2 \setminus A_1 \setminus A_3= A_2 \setminus \{y_1,y_2\} \, , \\
B_3 &= A_3 \setminus A_2= A_3 \setminus \{y_2\} \, .
\end{align*}
Consider any $i\in\{1,2,3\}$.
Using Lemma~\ref{lemma} as in Section~\ref{s:horizontal_BAs}, 
necessary conditions for a maximum likelihood estimator are that the ratio
$\frac{n(x)} {p(x)}$ is the same for all $x\in B_i$, namely
\begin{equation*}
\frac{n(x)} {p(x)}=
\frac{\sum\limits_{y\in B_i} n(y)} {\sum\limits_{y\in B_i} p(y)} =: 
\frac{n(B_i)} {p(B_i)}
=:K_i\,,
\end{equation*}
where we denoted 
\begin{align*}
n(B_i)&=\sum_{y\in B_i} n(y) \,,\\
p(B_i)&=\sum_{y\in B_i} p(y) \,.
\end{align*}
Equivalently,
\begin{equation*}
p(x|B_i) := 
\frac{p(x)} {p(B_i)}=
\frac{n(x)} {n(B_i)}\,.
\end{equation*}

\begin{figure}[ht]
\begin{center}
\unitlength=1mm
\special{em:linewidth 0.4pt}
\linethickness{0.4pt}
\begin{picture}(40,22)(0,15)
\put(30,30){\circle*{2}}
\put(10,30){\circle*{2}}
\put(10,20){\circle*{2}}
\put(20,30){\circle*{2}}
\put(30,20){\circle*{2}}
\put(30,20){\line(0,1){10}}
\put(30,30){\line(-1,0){20}}
\put(10,30){\line(0,-1){10}}
\put(32,18){\makebox(0,0)[lt]{$B_1$}}
\put(32,32){\makebox(0,0)[lb]{$y_2$}}
\put(8,32){\makebox(0,0)[rb]{$y_1$}}
\put(8,18){\makebox(0,0)[rt]{$B_3$}}
\put(20,32){\makebox(0,0)[cb]{$B_2$}}
\end{picture}
\caption{Simplified diagram of the second step of MLE of the diagram from Fig.~\protect\ref{f:Pi2}}
\label{f:simplified}
\end{center}
\end{figure}
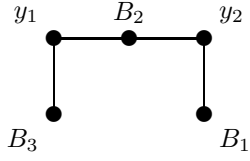

In the second step, the have to optimize the likelihood for a simplified structure which can be vaguely represented by Fig.~\ref{f:simplified};
here $B_i$ play a role of outcomes rather than sets of outcomes.
From the absolute frequencies
$n(B_1)$, $n(y_1)$, $n(B_2)$, $n(y_2)$, $n(B_3)$, 
we have to estimate the corresponding probabilities, bound by the restrictions
\begin{align*}
 p(B_1)+p(y_1) &=1\,,\\
 p(B_2)+p(y_1)+p(y_2)  &=1\,,\\
 p(B_3)+p(y_2) &=1\,.
\end{align*}
 
We 
define:
\begin{align*}
c_1 &:=\frac{K_2} {K_1+K_2}
= \frac{n(x_2) \, p(x_1)}{n(x_1) \, p(x_2) + n(x_2) \, p(x_1)} \, , \\
c_2 &:=\frac{K_2} {K_3+K_2}
= \frac{n(x_2) \, p(x_3)}{n(x_3) \, p(x_2) + n(x_2) \, p(x_3)} 
\, .
\end{align*}
 
The MLE solution of the diagram from Fig.~\ref{f:simplified}
gives conditions:
\begin{align*}
\frac{n(B_1)}{p(B_1)} &= n(B_1) + (1-c_1) \, n(y_1) \, , \\
\frac{n(B_2)}{p(B_2)} &= n(B_2) + c_1 \, n(y_1) + c_2 \, n(y_2) \, , \\
\frac{n(B_3)}{p(B_3)} &= n(B_3) + (1-c_2) \, n(y_2) \, , \\
\frac{n(y_i)}{p(y_i)} &= \frac{n(B_2) + c_1 \, n(y_1) + c_2 \, n(y_2)} {c_i} && \forall i \in \{1,2\} \,.
\end{align*} 
Then the MLE solution of Fig.~\ref{f:simplified} takes the following form:
\begin{align*}
p(x) &= \frac{n(x)}{n(B_1) + (1-c_1) \, n(y_1)} && \forall x \in B_1 \, , \\
p(x) &= \frac{n(x)}{n(B_2) + c_1 \, n(y_1) + c_2 \, n(y_2)} && \forall x \in B_2 \, , \\
p(x) &= \frac{n(x)}{n(B_3) + (1-c_2) \, n(y_2)} && \forall x \in B_3 \, , \\
p(y_i) &= \frac{c_i \, n(y_i)}{n(B_2) + c_1 \, n(y_1) + c_2 \, n(y_2)} && \forall i \in \{1,2\} \, .
\end{align*}
Elimination of
\begin{equation*}
c_1 = \frac{n(B_2) + c_2 \, n(y_2)}{n(B_1) + n(B_2) + c_2 \, n(y_2)} 
\end{equation*}
gives a quadratic equation for $c_2$, 
which has exactly one solution satisfying the assumptions.
 
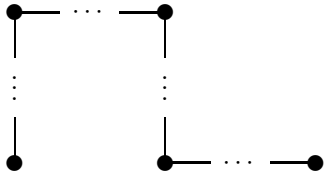
\begin{figure}[ht]
\begin{center}
\unitlength=1mm
\special{em:linewidth 0.4pt}
\linethickness{0.4pt}
\begin{picture}(60,32)(0,5)
\put(30,30){\circle*{2}}
\put(30,10){\circle*{2}}
\put(10,10){\circle*{2}}
\put(10,30){\circle*{2}}
\put(10,21){\makebox(0,0)[cc]{$\vdots$}}
\put(20,30){\makebox(0,0)[cc]{$\cdots$}}
\put(30,21){\makebox(0,0)[cc]{$\vdots$}}
\put(50,10){\circle*{2}}
\put(40,10){\makebox(0,0)[cc]{$\cdots$}}
\put(30,10){\line(1,0){6}}
\put(50,10){\line(-1,0){6}}
%
\put(30,10){\line(0,1){6}}
\put(30,30){\line(-1,0){6}}
\put(30,30){\line(0,-1){6}}
\put(10,30){\line(1,0){6}}
\put(10,30){\line(0,-1){6}}
\put(10,10){\line(0,1){6}}
\end{picture}
\caption{Greechie diagram of a non-constructible lattice for which MLE is feasible, with three splitting parameters}
\label{f:chain_short}
\end{center}
\end{figure}

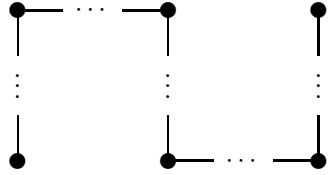
\begin{figure}[ht]
\begin{center}
\unitlength=1mm
\special{em:linewidth 0.4pt}
\linethickness{0.4pt}
\begin{picture}(60,32)(0,5)
\put(30,30){\circle*{2}}
\put(30,10){\circle*{2}}
\put(10,10){\circle*{2}}
\put(10,30){\circle*{2}}
\put(10,21){\makebox(0,0)[cc]{$\vdots$}}
\put(20,30){\makebox(0,0)[cc]{$\cdots$}}
\put(30,21){\makebox(0,0)[cc]{$\vdots$}}
\put(50,10){\circle*{2}}
\put(50,30){\circle*{2}}
\put(40,10){\makebox(0,0)[cc]{$\cdots$}}
\put(50,21){\makebox(0,0)[cc]{$\vdots$}}
\put(30,10){\line(1,0){6}}
\put(50,10){\line(-1,0){6}}
\put(50,10){\line(0,1){6}}
\put(50,30){\line(0,-1){6}}
\put(30,10){\line(0,1){6}}
\put(30,30){\line(-1,0){6}}
\put(30,30){\line(0,-1){6}}
\put(10,30){\line(1,0){6}}
\put(10,30){\line(0,-1){6}}
\put(10,10){\line(0,1){6}}
\end{picture}
\caption{Greechie diagram of a non-constructible orthomodular lattice for which MLE leads to four splitting parameters}
\label{f:chain}
\end{center}
\end{figure}

We can notice that this solution is identical to a MLE solution of horizontal sums where the occurrences of vertices contained within more than one operation were split between those operations by $c_1$ and $c_2$ respectively.
The same approach with different numbers of splitting parameters $(c_1, c_2, \ldots)$ could also be applied to Greechie diagrams from  Figs.~\ref{f:MO2x2}, \ref{f:Pi}, \ref{f:chain_short}, and~\ref{f:chain}.
 
This gives rise to new intuition about MLE on Greechie diagrams from Figs.~\ref{f:MO2x2}, \ref{f:Pi}, \ref{f:Pi2}, \ref{f:chain_short}, and~\ref{f:chain}.
We transform these Greechie diagrams to Greechie diagrams of horizontal sums by duplicating vertices found within multiple operations and splitting the recorded occurrences of the original vertices between them and their duplicates by newly introduced splitting parameters.
The MLE solution for horizontal sums is known, so all we have to do is to find the values of the aforementioned splitting parameters.
 
 For acyclic Greechie diagrams $(A,\B)$, where $\B=\bigcup_i \B_i$, $\B_i \cap \B_{i+1} = \{y_i\}$ for all but the last $\B_i$, $y_i\ne y_j$ 
 and $\B_i \cap \B_j = \emptyset$ for $i \ne j$ like in Figs.~\ref{f:MO2x2}, \ref{f:Pi}, \ref{f:Pi2}, \ref{f:chain_short}, and~\ref{f:chain}, the splitting parameters can be found as solutions to a polynomial of degree equal to the number of splitting parameters.
Only solutions on interval $[0,1]$ are admissible due to the way the splitting parameters are defined.
As we are maximizing a strictly concave function, there is exactly one solution in interval $[0,1]$.
(It is easy to show that maximum likelihood cannot be achieved at the boundary.)
The MLE can be found numerically by methods of convex optimization (applied to the negative likelihood).
We derived analytic closed-form solutions for two or three splitting parameters.
 

\begin{figure}[ht]
\begin{center}
\unitlength=1mm
\special{em:linewidth 0.4pt}
\linethickness{0.4pt}
\begin{picture}(40,32)(0,5)
\put(30,30){\circle*{2}}
\put(30,10){\circle*{2}}
\put(10,10){\circle*{2}}
\put(10,30){\circle*{2}}
\put(10,20){\circle*{2}}
\put(20,10){\circle*{2}}
\put(20,30){\circle*{2}}
\put(30,20){\circle*{2}}
\put(20,20){\circle*{2}}
\put(20,10){\line(0,1){20}}
\put(30,10){\line(0,1){20}}
\put(30,30){\line(-1,0){20}}
\put(10,30){\line(0,-1){20}}
\end{picture}
\caption{Greechie diagram of an orthomodular lattice for which MLE is more complicated}
\label{f:comb}
\end{center}
\end{figure}
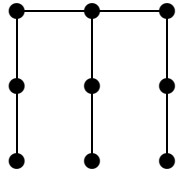

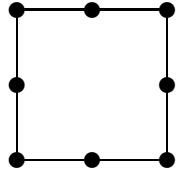
\begin{figure}[!ht]
\begin{center}
\unitlength=1mm
\special{em:linewidth 0.4pt}
\linethickness{0.4pt}
\begin{picture}(40,32)(0,5)
\put(30,30){\circle*{2}}
\put(30,10){\circle*{2}}
\put(10,10){\circle*{2}}
\put(10,30){\circle*{2}}
\put(10,20){\circle*{2}}
\put(20,10){\circle*{2}}
\put(20,30){\circle*{2}}
\put(30,20){\circle*{2}}
\put(10,10){\line(1,0){20}}
\put(30,10){\line(0,1){20}}
\put(30,30){\line(-1,0){20}}
\put(10,30){\line(0,-1){20}}
\end{picture}
\caption{Greechie diagram of an orthomodular poset for which MLE is more complicated}
\label{f:4cycle}
\end{center}
\end{figure}

There are more complex examples which are not of this kind; we leave them for future research.
The simplest two are in Fig.~\ref{f:comb} and \ref{f:4cycle} 
(which represents the smallest orthomodular poset which is not a lattice).

\section{Conclusion}

We collected arguments why maximum likelihood estimator (MLE) can be correctly defined for quantum probability (on orthomodular posets).
It admits a unique closed-form solution for a large class called constructible lattices.
Also for some other models MLE is feasible, although it leads to higher-order polynomial equations whose unique solution is not apparent.
For the general case, MLE exists and can be approximated numerically. 

\medskip
\textbf{Acknowledgments}
The first author was supported by the European Regional Development Fund, project ``Center for Advanced Applied Science'' (No.\ CZ.02.1.01/0.0/0.0/16\_019/0000778).
The second author received support from the Czech Science Foundation grant 20-09869L. 

%
%
%
%
%
%
%
%
%
%
%


\begin{thebibliography}{6}

\bibitem[\protect{Kl\"ay and Foulis}{1990}]{KlayFoulis}
{{Kl\"ay}, {M.P.}},
{{Foulis}, {D.J.}}:
{Maximum likelihood estimation on generalized sample spaces: An
  alternative resolution of {S}impson's paradox}.
{Found. Physics}
{20},
{777}--{799}
({1990})

\bibitem[\protect{Beran}{1984}]{Beran}
{{Beran}, {L.}}:
{Orthomodular Lattices. Algebraic Approach}.
{Academia},
{Prague}
({1984})

\bibitem[\protect{Kalmbach}{1983}]{Kalmbach}
{{Kalmbach}, {G.}}:
{Orthomodular Lattices}.
{Academic Press},
{London}
({1983})

\bibitem[\protect{Dichtl}{1981}]{Dichtl}
{{Dichtl}, {M.}}:
{Astroids and pastings}.
{Algebra Universalis}
{18},
{380}--{385}
({1981})

\bibitem[\protect{Greechie}{1971}]{Greechie}
{{Greechie}, {R.J.}}:
{Orthomodular lattices admitting no states}.
{J. Combin. Theory Ser. A}
{10},
{119}--{132}
({1971})

\bibitem[\protect{Janowitz}{1972}]{Janowitz}
{{Janowitz}, {M.F.}}:
{Constructible lattices}.
{J. Austr. Math. Soc.}
{14},
{311}--{316}
({1972})

\end{thebibliography}

\end{document}